# A millimeter-wave Bell Test using a ferrite parametric amplifier and a homodyne interferometer

Neil A. Salmon*,[1], Stephen R. Hoon[2]

Manchester Metropolitan University, All Saints, Oxford Road, Manchester, M15 6BH, UK.

ABSTRACT

A combined ferrite parametric amplifier and millimeter-wave homodyne interferometer are proposed as an ambient temperature Bell Test. It is shown that the non-linear magnetic susceptibility of the yttrium iron garnet (YIG) ferrite, on account of its narrow line-width Larmor precessional resonance, make it an ideal material for the creation of entangled photons. These can be measured using a homodyne interferometer, as the much larger number of thermally generated photons associated with ambient temperature emission can be screened out. The proposed architecture may enable YIG quantum technology-based sensors to be developed, mimicking in the millimeter-wave band the large number of quantum optical experiments in the near-infrared and visible regions which had been made possible by use of the nonlinear beta barium borate ferroelectric, an analogue of YIG. It is illustrated here how the YIG parametric amplifier can reproduce quantum optical Type I and Type II wave interactions, which can be used to create entangled photons in the millimeter-wave band. It is estimated that when half a cubic centimeter of YIG crystal is placed in a magnetic field of a few Tesla and pumped with 5 Watts of millimeter-wave radiation, approximately $0.5 \times 10^{12}$ entangled millimeter-wave photon pairs per second are generated by the spin-wave interaction. This means an integration time of only a few tens of seconds is needed for a successful Bell Test. A successful demonstration of this will lead to novel architectures of entanglement-based quantum technology room temperature sensors, re-envisioning YIG as a modern quantum material.

## 1. Introduction

The Einstein-Poldolsky-Rosen (EPR) paper [1] from 1935 provoked a dialogue questioning the nature of reality at the quantum level, which stimulated hugely the development of quantum theory. The ensuing discussions around this are likely to extend well into the 21st century, as they are sitting right at the heart of emerging quantum computers and sensors.

John Bell in 1964 conceived a simple test on the nature of local reality [2] that centered on evaluating an inequality. If the inequality is obeyed, future events in a system are completely determined by past history, this information being passed on in postulated hidden variables. The term hidden is used as these variables are not part of quantum theory. If the inequality is violated, any local hidden variables are not dictating future events, and the state of a system is determined only by its measurement.

John Clauser experimentally embodied [3] the Bell Test by using polarized optical photons; similar tests followed and these types of test are now referred to as Bell Tests. The maturing technologies of nonlinear dielectric crystals, lasers and photo-diodes then enabled the first demonstrations of a Bell Test that violated the Bell inequality by 40 standard deviations [4]. With the exceptions of a few loopholes, this demonstrated that at the quantum level, reality is defined by a measurement. Since then optical Bell Tests have closed many of the loopholes [5] and reduced signal integration times from 200 hours in the 1970's [6], to a few seconds with today's technology. This technology in the optical band is now available commercially for undergraduate teaching laboratories [7].

The initial Bell Tests in the optical band were optimal, as detection of the weak entangled photon flux went unhindered by spurious signals from thermally generated photons. For operation in the millimeter wave band the problem of the thermal photons has been successfully combated by cryogenically cooling the sources and receivers [8], [9].

For operation at ambient temperature in the millimeter wave band, a novel Bell Test is herein proposed. Such a system would be cheaper and lead to easier experimentation and in-field system deployments. The thermal photon problem is circumvented by using a yttrium iron garnet (YIG) ferrite parametric amplifier source of entangled photons with a phase link to a homodyne interferometer performing coherent integration [10]. A key point about this test is that the pump which generates the entangled photons is used to coherently integrate the entangled photon signature in quadrature space. It is important to recognize that the special phase relationship between the pump, signal and idler is a fundamental enabler of the coherent integration in the proposed Bell Test.

The proposed homodyne system generates non-degenerate entangled photons over a quasi-continuous spectrum, and then measures their phases and amplitudes, as proposed in [11] for the optical band. This type of measurement is referred to as one of continuous-variable (CV). The other type of entanglement measurement is that of the discrete-variable (DV), where only two energy states are involved. Successful Bell Tests (invalidating the Bell inequality) were first demonstrated using discrete-variable measurements. Successful continuous-variable Bell Tests came later [12].

The method section in this paper analyses the numbers of millimeter wave thermal photons produced at ambient temperature and explains how entangled photons could be generated using nonlinear magnetic materials such as a YIG ferrite. The results section determines the flux of entangled photon pairs in relation to practicable levels of parametric pump power. The discussion section describes the types of millimeter wave circuits that could be used to induce Type I and Type II interactions and how these may be used to generate entangled photons and the Bell States. Opportunities to close possible Bell Test loopholes are also discussed.

## 2. Method

Semiclassical considerations are now made as to the numbers of millimeter wave thermal photons that are generated at ambient temperature and how entangled millimeter wave photons may be generated using nonlinear susceptible materials, in particular YIG ferrites.

### 2.1. Thermally generated photons in the Rayleigh-Jeans and quantum sensing regimes

The ratio of photon to thermal energy ($hf/kT$) makes a clear distinction between the Rayleigh-Jeans regime ($hf/kT < 1$) and the quantum regime ($hf/kT>1$), where $f$ is the photon frequency, $T$ is ambient temperature, $h$ is Planck's constant and $k$ is Boltzmann's constant. At the ambient temperature (290 K) the photon energy becomes equal to the thermal energy ($hf=kT$) at 6 THz; below this frequency is the Rayleigh-Jean's regime and above this is the quantum regime.

Being either in the Rayleigh-Jeans regime or the quantum regime has a major effect on the mean number of thermally generated photons $\bar{n}$ which occupy a mode in the sensor. The mean number of photons is given [13] by the Bose-Einstein distribution function

$$\bar{n} = \frac{1}{e^{hf/kT} - 1} \qquad (1)$$

which at ambient temperature in the optical band (~$5 \times 10^{14}$ Hz) becomes $exp\ (-hf/kT)$, of the order of $2 \times 10^{-22}$ and in the microwave band (~10 GHz) becomes $kT/hf$, of the order of 604. This indicates the heart of the problem: In the millimeter wave band, the number of thermally generated photons is high and likely to swamp the relatively low fluxes of entangled photons.

* Corresponding Author.
   *E-mail address:* n.salmon@mmu.ac.uk (N.A. Salmon).
[1] ORCid: 0000-0003-4786-7130.
[2] ORCid: 0000-0002-1250-9432.

Because of the large numbers of thermally generated photons at ambient temperature the photon counting techniques used in the optical band are completely unsuitable in this frequency band. The above figure of 604 photons indicates there are these numbers of photons per unit bandwidth. This indicates if the classical radiation field were sampled at the Nyquist rate there would be this number of photons in a single sample. If there were one or more entangled photons also there, they would be completely swamped by the thermal photon flux. For this reason, the technique proposed here uses radio receivers to measuring in quadrature space, with a phase link to the parametric amplifier pump to enable coherent integration.

*2.2. Generation of entangled photons in a nonlinear medium*

Pair production is a process whereby an incident photon vanishes and two photons appear in its place, whose energies sum to that of the original photon. It is a quantum process happening across the electromagnetic spectrum, from microwaves to gamma rays. The incident photon must interact with matter for the process to proceed, as this provides energy levels with which the electromagnetic wave can interact and generate entangled photons. A Feynman diagram for the process is illustrated in Fig. 1.

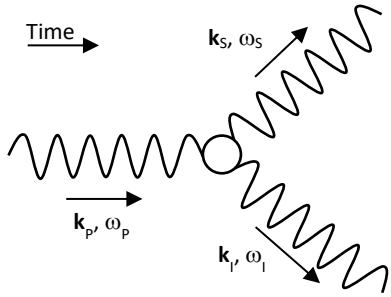

**Fig. 1.** Feynman diagram for SPDC pair production, where a pump (P) photon vanishes and a signal (S) and idler (I) photon appear in its place.

In quantum optics entangled photons are generated when the electric field **E** from a laser beam is sufficiently intense to force the dielectric susceptibility into a nonlinear region. The dielectric susceptibility $\chi_E$ is defined by Eq. (2) where $P$ is the material polarization [14].

$$P = \varepsilon_0 \left( \chi_{(1)E} E + \chi_{(2)E} : E^2 + \chi_{(3)E} :: E^3 + \cdots + \chi_{(n)E} : \dots : E^n \right) \quad (2)$$

At low intensities only the linear susceptibility $\chi_{(1)E}$ is active, but at higher fields the $n^{th}$ order nonlinear susceptibilities $\chi_{(n\geq 2)E}$ become active, this happening more readily in non-centrosymmetric crystals. The unit cells of these crystals can have permanent dipole electric moments and these may spontaneously align to form a domain. These types of materials are referred to as pyro- or ferro-electric and are used to generate visible entangled photons. The most commonly used material for this at present is beta barium borate (BBO).

Ferromagnetic materials are the magnetic equivalent of ferroelectric ones, as magnetic moments in the unit cells spontaneously align to form a domain. Upon the application of a magnetizing intensity **H,** a magnetization **M** results, the relationship being highly nonlinear and hysteretic, and described by

$$\mathbf{M} = \chi_{(1)M} \mathbf{H} + \chi_{(2)M} : \mathbf{H}^2 + \chi_{(3)M} :: \mathbf{H}^3 + \cdots + \chi_{(n)M} : \dots : \mathbf{H}^n \quad (3)$$

where $\chi_{(1)M}$ is the linear magnetic susceptibility and $\chi_{(n\geq 2)M}$ are the higher order nonlinear susceptibilities [15]. These materials are not currently recognized as being potential sources of entangled photons.

Entangled photon generation takes place by a process known as spontaneous parametric fluorescence or spontaneous parametric down-conversion (SPDC). It is the second order (n=2) nonlinear susceptibility which is responsible for this three-wave interaction that creates a pair of entangled photons from a pump photon [16], the first experimental proof of this being presented in [17].

The semi-classical explanation for this process is that the nonlinear medium generates a photon that has a difference frequency, this being defined as the difference between the pump photon and a vacuum photon from the quantum zero-point field energy. For the difference frequency photons to be generated it is necessary for the second order nonlinear susceptibility $\chi_{(2)E}$ to be relatively large, typically in the region of a few pm/Volt for many ferroelectrics when excited by laser radiation.

*2.3. Energy and momentum conservation in entangled photon generation*

Conservation of energy dictates that the vacuum photon must have an energy less than that of the pump photon. The energy from the pump photon then passes to that of the signal and idler photons. Given the Einstein energy relation $E = \hbar\omega$, conservation of energy relates the frequencies of the pump $\omega_P$, signal $\omega_S$ and idler $\omega_I$ as

$$\omega_P = \omega_S + \omega_I. \quad (4)$$

From photon ballistics, the process must also satisfy the conservation of momentum. Given the de Broglie momentum relation $\mathbf{p} = \hbar\mathbf{k}$, where **k** is the direction propagation vector, conservation of momentum relates the propagation vectors of the pump $\mathbf{k}_P$, signal $\mathbf{k}_S$ and idler $\mathbf{K}_I$ as,

$$\mathbf{k}_P = \mathbf{k}_S + \mathbf{k}_I, \quad (5)$$

and in the vector triangle of Fig. 2, the magnitude of the propagation wave vector is related to the wavelength λ of the photon as $|\mathbf{k}|=2\pi/\lambda$.

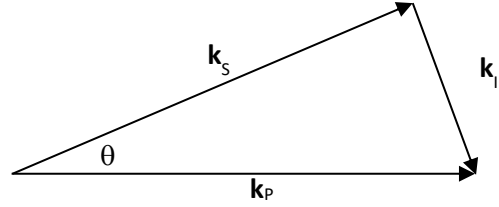

**Fig. 2.** The momentum (**k**) conservation vector triangle is shown for entangled photon pair creation, where θ is the angle between the pump and signal propagation vector.

Since the magnitude of the wave vector is related to its phase $\phi$, through the wave description $exp[j(\mathbf{k}.\mathbf{r} - \omega t + \phi)]$, the conservation of momentum indicates the phases of the pump, signal and idler at the point of pair creation are related as

$$\phi_P = \phi_S + \phi_I. \quad (6)$$

This phase relationship enables coherent integration to recover the signature of the signal and idler photons in the radio Bell Test from the high level of thermal noise characteristic of the Rayleigh-Jeans region. Without this basic and very fundamental phase relationship, no coherent integration would be possible.

*2.4. Phase matching to maximize the flux of entangled photon pairs*

Most materials display properties of *normal dispersion*, where refractive index rises with frequency. The refractive index usually varies with the direction of propagation and it is often not possible to satisfy completely the conservation of momentum of Eq. (5) depicted in Fig. 2. In these cases, there is a mismatch in the wave-vector **k** and this is quantified as

$$\Delta\mathbf{k} = \mathbf{k}_P - \mathbf{k}_S - \mathbf{k}_I. \quad (7)$$

Since the magnitude of the **k**-vector can be written $\omega\mathbf{n}/c$, where n is the refractive index in the direction of propagation, the above k-vector mismatch can be written as

$$\Delta\mathbf{k} = (\omega_P \mathbf{n}_P - \omega_S \mathbf{n}_S - \omega_I \mathbf{n}_I) \, 1/c. \quad (8)$$

Since matching the wave vectors is associated with the wavelengths of the pump, signal and idler radiation, achieving momentum conservation can be achieved by tailoring the refractive index and angles of propagation. This is referred to as phase matching. In cases where phase matching is achieved at an angle θ=0, it is referred to as collinear phase matching. The other cases, where θ ≠ 0, are referred to as non-collinear phase matching.

In the 1980's attention to effective phase matching increased the efficiency of entangled photon generation by several orders of magnitude. As a result, for a single crystal the probability of detecting an entangled pair per pump photon is now in the region of $3\times10^{-11}$ [18]. Periodic poling in lithium niobate has increased this probability to $4\times10^{-6}$ per pump photon [19]. Other methods of generating entangled photons in the visible band have been SPDC in an AlGaAs semiconductor waveguide [20]. In the microwave band operating at cryogenic temperatures an entangled photon power of -110 dBm has been generated by pumping a Josephson junction (acting as a parametric amplifier) with -75.6 dBm, indicating efficiency of the order of -34.4 dB ($3.6 \times 10^{-4}$) [9].

Two relatively efficient interactions for generating entangled photons are referred to as Type I and Type II [14]. In a Type I interaction each photon of the entangled pair has the same (linear) polarization, which is orthogonal to the pump polarization. This is referred to as an 'e-oo' or an 'o-ee' type interaction, where 'o' refers to the ordinary or o-mode polarization (E-vector perpendicular to the crystal incidence plane) and 'e' refers to the extraordinary or e-mode polarization (E-vector parallel to the crystal incidence plane) [14], the crystal incidence plane being defined as that which contains the wave propagation vector **k** and the crystal optic axis. In the designation 'e-oo', the first letter 'e' refers to the polarization of the pump and the other two letters refer to the polarizations of the signal and idler photons.

In a Type II interaction, the polarizations of each photon of a pair are orthogonal to one another, with one of these polarizations being identical to that of the pump polarization, these interactions designated as 'e-oe', 'e-eo', 'o-oe' or 'o-eo'.

*2.5. Ferrites as a nonlinear medium for entangled photon generation*

The potential of ferrites as a nonlinear medium for the generation of entangled photons can be appreciated from the nonlinear dependence of the magnetization **M** on the static magnetizing intensity **H** shown in Fig. 3. Ferrites, a well understood class of materials [21], [22], can have their constitutive magnetic properties (eg. coercivity $H_C$, remanence $M_R$, saturation magnetization $M_S$, susceptibility $\chi$) finely tuned by appropriate doping and prepared in single crystal or thin film form. This is illustrated in Fig. 3 by the hysteresis loops of Ho doped YIG, described by $Y_{3-x}Ho_xFe_5O_{12}$, where x=0 and 1.5 [23], the curves being modelled heuristically using a modified form of the Langevin function $L(x)$, where $M = M_s(\coth \mu_0 a(H-H_c) - 1/\mu_0 a(H-H_c))$ and $a$ is a fitting parameter, from data presented in [23].

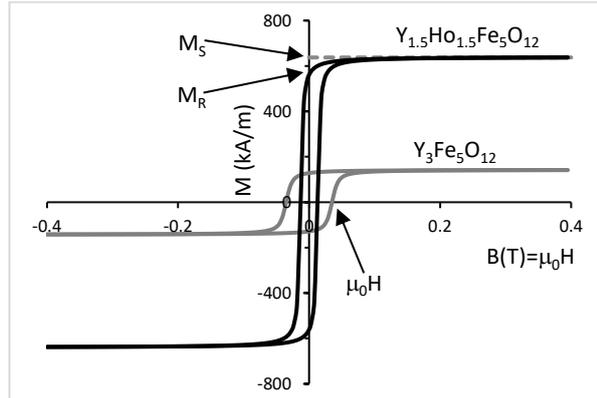

**Fig. 3.** The nonlinear magnetic susceptibility (M/H) evident from the hysteresis loop shown here is typical of a pure and Ho doped (x=1.5) YIG ferrite illustrating the high field (1T) saturation magnetization ($M_{S\ x=1.5}$ 0.804 T (i.e. 640 kA/m)) (remanent magnetization ($M_{R\ x=1.5}$ =561 A/m) and coercivity ($H_{C\ x=1.5}$ = 0.013 T).

At the atomic level in ferrites, the Heisenberg uncertainty principle prevents the net atomic or domain magnetic moment (related to angular momentum) aligning with the magnetizing intensity vector **H**, as angular momentum and its direction don't commute. However, the torque to align induces a precessional motion of the magnetic moment around the field direction at the (classical) Larmor precession frequency [24] of

$$\omega_L = -\gamma_S H ,  \qquad (9)$$

where

$$\gamma_S = -g_e \frac{\mu_0 e}{2m_e} . \qquad (10)$$

is the electron spin gyro magnetic ratio and *e* and $m_e$ are the electron charge and mass respectively, and $g_e$ is the electron Landé g-factor, having a value ~2.002319. The Larmor precession frequency is equivalent to the frequency of the photons absorbed in the quantum mechanical representation as magnetic moments flip in an applied field during resonant absorption between spin up and spin down states [25].

As the conduction bands in ferrites contain no free electrons their resistivity is high, typically $10^7 \Omega m$, so the material is transparent to microwaves and millimeter waves. This means the ferrite interaction of radiation is governed by the complex refractive index given [26] by

$$n = \sqrt{\varepsilon_r \mu_r} , \qquad (11)$$

where $\varepsilon_r$ (=$\varepsilon_r$' - j$\varepsilon_r$'') and $\mu_r$ (=$\mu_r$' - j$\mu_r$'')) are the complex relative permittivity and permeability of the material, these being related to the first order (linear) susceptibility as $\varepsilon_r = 1 + \chi_{(1)E}$. For ferrites the complex relative permittivity ($\varepsilon_r$', j$\varepsilon_r$'') is [27] isotropic and takes a scalar value, typically (12-22, 0.05). The permeability of the ferrite is dictated by the Polder Tensor [24] and the variation of this with frequency over the millimeter wave band determines largely how these materials behave and have been exploited as devices in a large commercial market.

The ferrite response to electromagnetic radiation falls into two orthogonal propagation modes, one coupled strongly to the magnetic dipoles and the other weakly coupled, the strength being dependent on the angle of propagation in the medium. For longitudinal propagation (along the static magnetic field lines) the strongly coupled mode is right-hand circularly (RHC) polarized and the weakly coupled mode is left-hand circularly (LHC) polarized. For transverse propagation (at right angles to the static field lines) the strongly coupled mode is the e-mode (the H-field of the wave being perpendicular to the static magnetic field direction), with the weakly coupled mode being the o-mode (having the wave H-field parallel to the static field direction). For the strong mode the ferromagnetic resonance shifts from $\omega_0=\omega_L$ for longitudinal propagation to $\sqrt{(\omega_0(\omega_0+\omega_M))}$ where $\omega_M=-\gamma_S M$, for transverse propagation. At intermediate propagation angles the strong and weak modes are elliptically polarized.

Garnets are a class of ferrites of particular interest as they have very low damping and hence little crystal lattice absorption of the wave energy. This means the ferromagnetic resonance line is very narrow and in yttrium iron garnet (YIG) it has a width [28] of ~1 MHz, which gives it a highly non-linear magnetic susceptibility.

A plot of the complex refractive index is shown in Fig. 4 for transverse propagation in an infinite solid of YIG for a Larmor precession frequency (associated with the internal field H) of 15 GHz and a Larmor precession frequency (associated with the internal magnetization $\omega_M/2\pi$) of 6.9 GHz. The dielectric constant ($\varepsilon$') is assumed to be 14.7, the dielectric loss tan ($\varepsilon$''/$\varepsilon$') to be 0.0002 [29] and the damping factor $\alpha$ to be (~1/Q) of 0.00007 (typical for YIG) [30]. For the strongly coupled mode there is a (negative permeability) non-propagating region extending from the resonant frequency to the cut-off frequency, $\omega_0+\omega_M$, above which the permeability becomes positive.

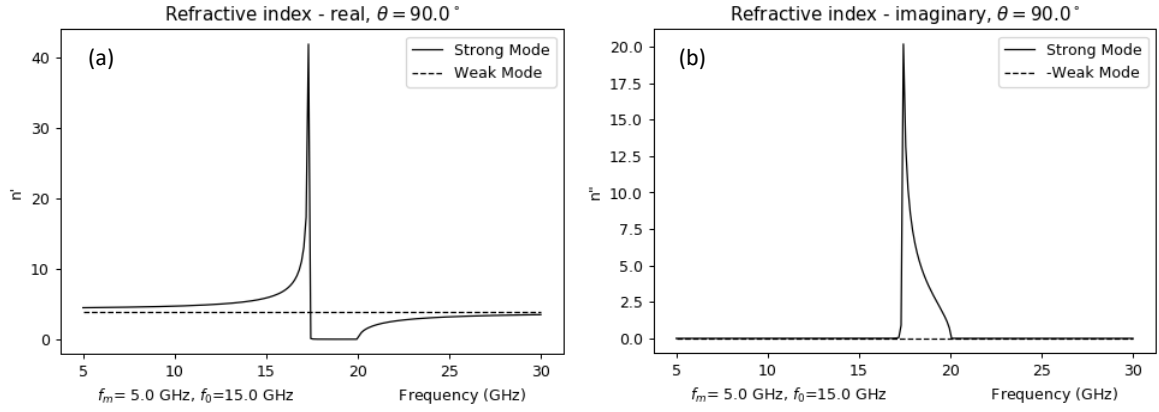

**Fig. 4.** The real (a) and imaginary (b) parts of the refractive index for the strong and weak transverse propagation modes in YIG.

In general the refractive index is a function of the propagation angle in the YIG and its quantitative behavior is well known from the solution to Eq. (11), as illustrated to good effect in [24]. From this equation two orthogonal elliptical polarization modes arise (which become linearly (circularly) polarized for propagation across (along) the magnetic field lines. This behavior is similar to the angular variation in the refractive index for the (extra)ordinary mode in electro-optic (ferroelectric) crystals, as can be appreciated in the refractive index polar plots in [14]. Therefore there is an opportunity to use the angular variation of refractive index in YIG for phase matching (by minimizing the momentum mismatch $\Delta\mathbf{k}$ from Eq. (8)) to maximize the intensity of the entangled photons in accordance with Eq. (17). It may also be possible to increase the nonlinear interaction length in YIG (whilst retaining good phase matching), to increase the efficiency of entangled photon generation, by periodically poling YIG (PPYIG), as is done in periodically polled lithium niobate (PPLN) [31].

## 3. Results

Given the above mechanism for creating entangled photons, estimations are now made of the likely fluxes of entangled photons from YIG in the millimeter-wave band using moderate pump powers that are readily available. This will determine the practicability of the proposed Bell Test in this spectral band and therefore the likelihood that the system could be used as the basis for a sensor.

*3.1. The second order nonlinear magnetic susceptibility of YIG*

Optimization of the second order nonlinear susceptibility of ferrites has led to these devices being used as frequency doublers and parametric amplifiers [15]. The second order nonlinear susceptibility of a frequency doubling ferrite is given [32] as

$$\chi_{(2)M} = \frac{\gamma_S M_0}{\omega_L \Delta H}, \quad (12)$$

where $M_0$ is the YIG static magnetization, $\omega_L$ is the Larmor angular precession frequency associated with the externally applied static field $H_0$ (given by $-\gamma_S H_0$) and $\Delta H$ is the ferromagnetic resonance linewidth in A/m. It can be seen from this equation that a higher second order nonlinear susceptibility results from a narrow resonance width. Small resonance widths of the order of 0.36 Oersted (~28 A/m) can be achieved using a YIG crystal [32]. For a garnet with a static magnetization of 238 kA (~0.3 Tesla) pumped at a frequency of 20 GHz (to resonate with the externally applied field) the second order nonlinear susceptibility from Eq. (12) is ~0.015 m/A.

For frequency doubling a nonlinear response is achieved [32] by pumping in the transverse mode, with the B-field of the pump radiation being perpendicular to the static field. The frequency doubled component is also taken off in the transverse propagation configuration, with the B-field parallel to the static field.

For parametric amplification in YIG a number of modes are [15], [33] possible. These are analogous to the different types of optical entangled photon creation: just how these are excited is explained in the following section on the circuits for millimeter wave entangled photon creation. The mode with the highest gain is the magneto-static mode (or the spin-wave or three-magnon mode) [34]. Here a linearly polarized pump wave at twice the resonance frequency propagates transversely across the static magnetic field lines with its wave H-field parallel to the static field direction, almost an exact reverse process of the above frequency doubling mechanism. This system creates parametric gain in ferrites at pump power levels of 0.5 W, generating signal and idler photons that are linearly polarized with their H-fields perpendicular to the static field directions. Considering the direction of magnetization in a ferrite to be analogous to the optic axis in a ferroelectric, the mode is the magnetic equivalent of the Type I interaction in ferroelectrics.

A further parametric amplification mode is the electromagnetic mode. Here a linearly polarized pump wave at the ferromagnetic resonance frequency propagates transversely across the static magnetic field lines with its H-field perpendicular to the static field direction [35]. Signal and idler photons are created having orthogonal linear polarizations, one having the H-field perpendicular and the other with the H-field parallel to the static field direction. This mode is the magnetic equivalent of the Type II entangled photon creation in ferroelectrics.

Miller's rule [36] is an empirical relationship from the field of nonlinear optics for non-centrosymmetric dielectrics stating that the ratio of the second order nonlinear susceptibility to the product of the first order susceptibilities at the three frequencies ($\omega_1 + \omega_2, \omega_1, \omega_2$) of a three-wave interaction, namely

$$\frac{\chi_{(2)E}(\omega_1 + \omega_2, \omega_1, \omega_2)}{\chi_{(1)E}(\omega_1 + \omega_2)\chi_{(1)E}(\omega_1)\chi_{(1)E}(\omega_2)} \quad (13)$$

is constant.

Boyd shows [31] that the mathematical form of Miller's rule is a consequence of the anharmonic motion of the electron in the E-field of a passing electromagnetic wave, this anharmonicity being a direct result of the local crystal E-field in the non-centrosymmetric material. However, there is only finite second order nonlinear electrical susceptibility $\chi_{(2)E}$ if the crystal is non-centrosymmetric. This suggests that if YIG has a finite magnetic susceptibility $\chi_{(2)M}$ it should also follow Miller's rule.

Assuming Miller's rule is followed for the magnetic susceptibilities, this suggests that the nonlinear susceptibility of Eq. (12) is very close to that for difference frequency generation, provided the first order susceptibilities are similar. The value from Eq. (12) of 0.0015 m/A may therefore be used in the estimation of the nonlinear susceptibility for the difference frequency generation in spontaneous parametric down-conversion.

*3.2. The power of entangled photons*

In the process of entangled photon generation, the pump photon interacts with a vacuum photon to annihilate itself and create a pair of entangled photons. A high pump power parametrically amplifies [37] the vacuum photons. If the field-gain of the nonlinear medium is designated $\gamma$ and its length $l$, the spectral radiance of the entangled photon flux (in W/m²/sr/(radians/s) is given [38] by

$$I_{\omega\Omega}(\mathbf{k}) = I_{\omega\Omega}^{VAV}\frac{sinh^2[(\gamma^2 - |\Delta\mathbf{k}|^2/4)^{1/2}l]}{(1 - |\Delta\mathbf{k}|^2/4\gamma^2)}, \quad (14)$$

where $\Delta\mathbf{k}$ is the wavevector (phase) mismatch of Eq. (8), and $I_{\omega\Omega}^{VAV}$ is the spectral radiance of the vacuum photon fluctuations, given by

$$I_{\omega\Omega}^{VAV} = \frac{\hbar\omega^3 n_S^2}{8\pi^3 c^2}, \quad (15)$$

where $n_S$ is the medium refractive index. The vacuum photon fluctuations correspond to a single photon in each mode. The field-gain from [38] is given as

$$\gamma_E(m^{-1}) = \left(\frac{8\omega_S\omega_I I_P \mu_0}{c n_P}\right)^{\frac{1}{2}} \pi \chi_{(2)E}, \quad (16)$$

where $n_P$ is the medium refractive index for the pump radiation, which has an intensity (or irradiance) in W/m² given as $I_P$. Taking a pump power of $10^4$ W/m² (1 W/cm²), a nonlinear electrical susceptibility of $5 \times 10^{-12}$ m/V and a refractive index of 2.2, the field gain of the medium is $\sim 1.2 \times 10^{-5}$ per meter for a signal and idler frequency of 10 GHz. For a low gain medium in which $\gamma \ll \Delta\mathbf{k}$, Eq. (14) becomes

$$I_{\omega\Omega}(\mathbf{k}) = I_{\omega\Omega}^{VAV} \gamma^2 l^2 sinc^2[(|\Delta\mathbf{k}|)l/2] \quad (17)$$

and if perfect phase matching is satisfied the entangled photon spectral radiance in the nonlinear dielectric medium becomes

$$I_{\omega\Omega}(\mathbf{k}) = \frac{\hbar \mu_0 \omega_S^4 \omega_I I_P n_S^2 \chi_{(2)E}^2 l^2}{\pi c^3 n_P}. \quad (18)$$

This indicates that for the above pump power the spectral irradiance of the entangled photon flux would be $6.88 \times 10^{-32}$ Watts/sr/m²/(radians/s) for an interaction length of 10 cm. If it is assumed the bandwidth is 5 GHz and emission is measured over $\pi$ steradians, the entangled power flux is $6.79 \times 10^{-25}$ Watts over an area of one square centimeter.

Making a similar analysis to [38] for entangled photon generation from a magnetic material the field-gain is given by

$$\gamma_M(m^{-1}) = \left(\frac{8\omega_S \omega_I I_P \mu_0}{cn_P}\right)^{\frac{1}{2}} \frac{\pi \chi_{(2)M}}{Z_P}, \quad (19)$$

where $\chi_{(2)M}$ is the 2nd order nonlinear magnetic susceptibility and $Z_P$ is the medium impedance to the pump radiation $Z_P = \sqrt{\mu_r \mu_0 / \varepsilon_r \varepsilon_0}$. Pumping YIG possessing a nonlinear magnetic susceptibility of 0.015 m/A and a refractive index of 3.8 with a pump power of $10^4$ W/m² (1 W/cm²) at 20 GHz results in a field gain from Eq. (19) of 630 per meter. This is a considerably higher gain than that of a nonlinear dielectric. Assuming that the conditions of phase matching are satisfied (ie $\Delta\mathbf{k}=0$), then the entangled photon spectral radiance in the nonlinear magnetic medium from Eq. (14) becomes

$$I_{\omega\Omega}(\mathbf{k}) = \frac{\hbar \omega^3 n_S^2}{8\pi^3 c^2} \sinh^2(\gamma l). \quad (20)$$

Given the estimated magnetic nonlinear susceptibility of 0.015 m/A from above and a refractive index of 3.8 for YIG, the entangled photon spectral irradiance becomes $1.80 \times 10^{-19}$ Watts/sr/m²/(radians/s) for an interaction length of 3 mm at signal and idler frequencies of 10 GHz. Assuming a bandwidth of 10 GHz and that emission is collected over $\pi$ steradians, the entangled photon flux for a pump power of 5 W/cm² is $3.56 \times 10^{-12}$ Watts/cm², which is approximately $0.5 \times 10^{12}$ photons per second from an area of 1 cm². This is also considerably higher than the above counterpart for a nonlinear dielectric medium.

YIG ferrites have many potential advantages when it comes to the generation of entangled states. YIG, as a bulk crystal, regular prism or thin film, can be immersed in an applied static bias field, $H_0$, close to the resonance radio frequency field, mounted in either a waveguide [39] or a very high quality factor (Q~5000) resonant cavity [40]. The high Q-factor raises the magnetizing intensity of the applied radiation, increasing the probability of generating entangled photons.

## 4. Discussion

Given the results on the fluxes of millimeter wave entangled photons, a discussion is opened here on architectures for their generation and a homodyne receiver system for their detection.

### 4.1. Circuits for Type I & Type II ferrite sources of entangled photons

Given that the cross-sectional area of the YIG crystal is envisaged to be in the region of 1 cm², the pump beam would need to be either focused from a waveguide or transmitted direct to the sample using a transmission line. Because the sample cross-sectional dimension is of the order of the wavelength of the signal and idler photons, these will emerge from the YIG crystal into a wide angled cone due to diffraction. This cone would constitute a single mode of the receiver system so signal and idler may be captured by the receiver regardless of energy. This represents an essential difference between what is proposed here and the quantum optics experiments in entanglement.

A circuit using the magneto-static mode of a parametric amplifier to create entangled photons is illustrated in Fig. 5. This exploits the Type I interaction and is similar to that in [41], but differs in that it uses a resonant circuit. The system comprises a pump generating linearly polarized radiation orientated at an angle of 45°, which splits into two orthogonal horizontally and vertically polarized beams using a vertically wired grid polarizer [42]. The horizontally polarized (E-field) beam enters the YIG-1 sample with the radio-frequency H-field parallel to the static H-field in the ferrite. Signal and idler photons generated with H-fields perpendicular to the static H-field exit via a reflection from the second vertically wired grid. The vertically polarized (E-field) beam enters the YIG-2 sample with the H-field perpendicular to the static H-field stimulating signal and idler photons which exit by transmission through the second wire grid.

The unused horizontal and vertically polarized pump radiation exit via the second wire grid polarizer with a relative phase (epoch) angle defined by the difference in phase delays between the two arms of the system. An adjustable wave-plate then provides sufficient delay between these polarizations to generate linear polarization orientated at -45°. This maximizes the amount of radiation that reenters the cavity by reflection from the 45°-angled polarizer. The polarization rotator is then adjusted to bring the orientation around to +45°, in order that equal fluxes of horizontal and vertical polarizations can be presented to the two YIG ferrite crystals. The entangled photons exit through the vertical wire grid polarizing beam splitter and are then separated into two channels by the following beam splitter so they may enter the homodyne interferometer of Fig. 8.

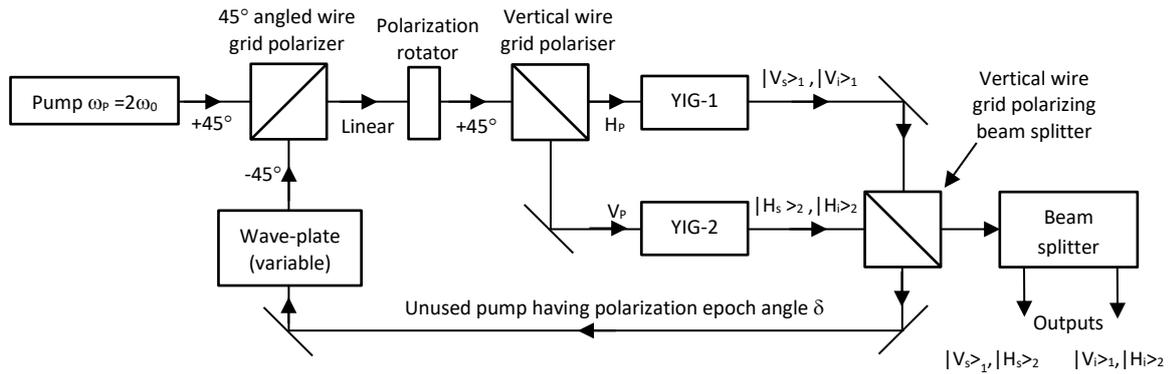

**Fig. 5.** Type I interaction in a resonant circuit using the magneto-static parametric amplifier mode in a YIG ferrite pumping at twice the ferromagnetic resonance frequency generates signal and idler photons about the ferromagnetic resonance frequency $\omega_0$ and the entangled state $(|V_S>_1|V_i>_1+e^{j\phi}|H_S>_2|H_i>_2)/\sqrt{2}$. The orientations of linear polarizations are indicated as +45°, -45°, H and V.

The beam splitter of Fig. 5 from which the signal and idler emerge can be realized in practice in several different forms. 1) It may be a non-polarizing 50/50 beam splitter, similar to [43]. 2) It may be a high/low pass filter, passing photons of frequency $> \omega_P/2$ into one channel and those $< \omega_P/2$ into the other channel. 3) It may be a polarizing grid orientated at 45°. The latter two forms would however make a preselection of photons on the basis of their momentum or polarization and thereby may remove one of the degrees of entanglement. However, since signal and idler are initially entangled in both momentum and polarization, removing one of these may still enable a Bell Test to be constructed.

A circuit using the electromagnetic mode of a parametric amplifier, which exploits the Type II interaction for the creation of entangled photons, is shown in Fig. 6. Pumping the YIG at the ferromagnetic resonance transversely with the H-field perpendicular to the static magnetic field direction, the signal and idler can be taken off transversely, one with the wave H-field perpendicular to the static field and the other parallel to the static field. The signal and idler are taken off using a dichroic mirror fabricated from a mesh grid [42]. The unused radiation gets reflected by the mesh filter and as the pump is horizontally polarized it passes through a half-wave plate to become vertically polarized, enabling it to re-enter the resonance cavity by reflection off the gird. The polarization rotator rotates the plane of polarization so that horizontally polarized radiation is incident on the YIG.

On exit via the dichroic mirror the entangled photons pass through a beam splitter, similar to that in the Type I system above.

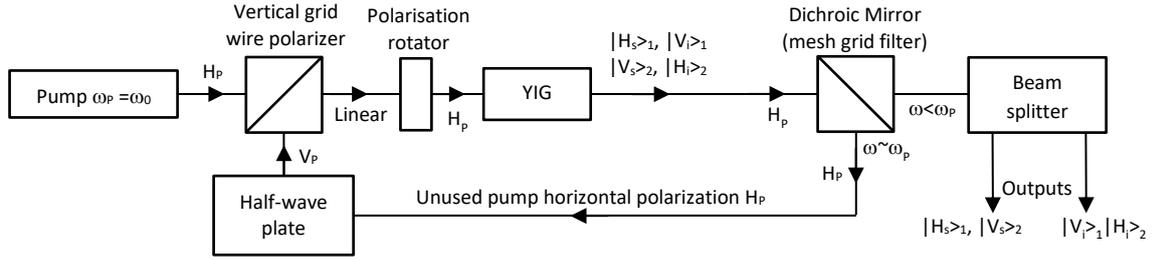

**Fig. 6.** Type II interaction in a resonant circuit using the electromagnetic mode parametric amplifier in YIG pumping at the ferromagnetic resonance frequency generating signal and idler photons about half the ferromagnetic resonance frequency, creating the entangled state $(|H_s\rangle_1|V_i\rangle_1+e^{j\varphi}|V_s\rangle_2|H_i\rangle_2)/\sqrt{2}$. The orientations of the linear polarizations throughout the system are indicated as H and V.

A non-resonant circuit to stimulate Type II interactions is adopted from the Sagnac interferometer of [44] and depicted in Fig. 7. Here linearly polarized pump radiation orientated at 45° encounters a polarizing beam-splitter constructed from a vertical grid. A horizontal polarization component of the pump passes through the grid and is converted to vertical polarization by the half wave-plate, before stimulating the YIG to generate entangled photons in mode 1. A vertical polarization component is reflected from the grid and stimulates the YIG crystal directly from the other direction to create entangled photons in mode 2. The Type II interactions create orthogonal signal and idler photons in these two modes which exit via the polarizing beam-splitter and the dichroic mirror. The dichroic mirror, manufactured from mesh grids [42], is designed to pass the pump radiation, but reflect the lower frequency signal and idler radiations. The outputs from the circuit pass directly into the channels of the homodyne interferometer of Fig. 8. The advantage of this architecture is that the photons remain entangled in both momentum and polarization. The unused pump radiation from the Sagnac system exits in the direction of the pump, but this could re-enter the system by a reconfiguration using additional resonant circuit components, not shown here.

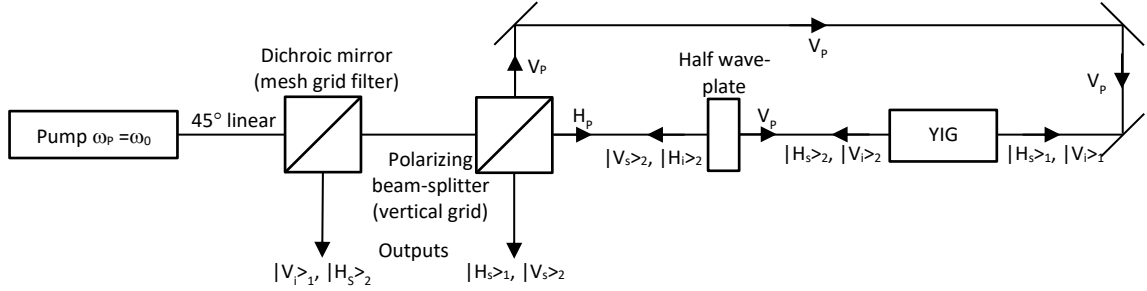

**Fig. 7.** A circuit based on a Sagnac interferometer [44] using a Type II interaction of the electromagnetic mode in YIG to create the entangled state $(|V_i\rangle_1|H_s\rangle_1 + e^{j\varphi}|H_s\rangle_2|V_s\rangle_2)/\sqrt{2}$.

*4.2. The homodyne interferometer Bell Test*

A single-channel Bell Test architecture, the type initially proposed in [3], but exploiting the attributes of millimeter wave receivers for an ambient temperature Bell Test is shown in Fig. 8. It creates the entangled photons by pumping a nonlinear YIG crystal, in the circuits of Fig. 5 (for entangled photons created by a Type I interaction) and Fig. 6 or Fig. 7 (for entangled photons created by a Type II interaction), and then separates the pair into two channels.

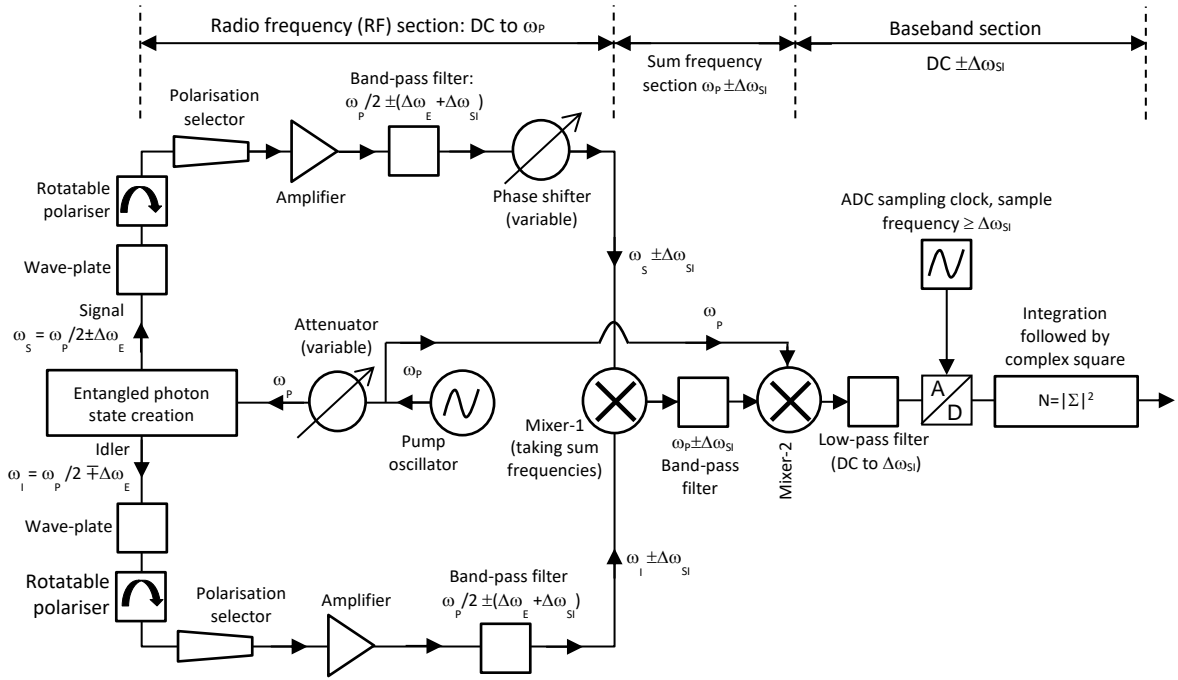

**Fig. 8.** The homodyne interferometer single-channel Bell Test is proposed to overcome the noise generated by thermal photons when operating in the Rayleigh-Jeans regime ($hf<kT$) [10], where $2\Delta\omega_E$ is the angular frequency difference between the signal and idler photons and $2\Delta\omega_{SI}$ is the assumed angular bandwidth of the signal and idler photons. The box labelled 'entangled photon state creation' represents any of the circuits from Fig. 5, Fig. 6 or Fig. 7.

Wave-plates are included in the circuit of Fig. 8 so that all Bell States can be created, as indicated in [45]. Since the system is an interferometer it is sensitive to phase, so adjustments to the phase can be made by the variable phase shifter in one of the arms of the interferometer. The effects of birefringence in components of the system could potentially lead to a reduced level of entanglement, but these may be compensated for by extra components in the circuity of Fig. 8. Since signal and idler photons need to arrive at more or less the same time at the receivers, the transit time of photons could be balanced by physical movement or by insertion of low-loss refractive index material into one of the arms of the interferometer, to increase its effective optical path length. The above adjustments would be made to maximize the measured entanglement signature, similar to that in the optical band described in [46].

The radiation in each channel is amplified to drive mixer-1 into its nonlinear regime. The mixer output contains the multiplication of signal and idler fields together with relatively large amounts of thermal noise on a 20 GHz carrier frequency. The phases of the signal and idler are summed by this mixing process. The output from mixer-1 is then mixed with the original pump radiation at 20 GHz in mixer-2, thereby shifting the combined signal and idler signature down to baseband.

The mixer-2 output is lowpass filtered and sampled in a sine and cosine receiver to record a complex amplitude, which is then coherently integrated to recover the stationary complex value for the signature of the signal and idler photon pairs from the noise. The sine and cosine receiver is also referred to by some as an I,Q or a quadrature receiver. It integrates to a stationary complex value because the phase noise $\phi_n$ averages to zero, leaving the phase relation of Eq. (6) to prevail. The square of this stationary value is equivalent to the number of coincidence counts, usually designated as $N$, in the optical Bell Test, as indicated by Eq. (21).

$$N = A^2 \Leftarrow |Ae^{j(\phi_s+\phi_I)}|^2 \Leftarrow |<A_n e^{j(\phi_s+\phi_I+\phi_n)}>|^2 \quad (21)$$

The circuit in Fig. 8 is the radio receiver equivalent of a single channel Bell Test. The Bell Test test-statistic, usually designated as $S$, is formed by the usual combination of measurements of $N$ at the Bell Test angles [3]. Measurements made at the Bell Test angles maximize the difference between the classical and quantum results, this giving the highest chance of demonstrating violation of the Bell inequality.

A twin-channel Bell Test architecture, also proposed in [3], but suitable for ambient temperature operation in the millimeter wave band, is constructed by replacing the polarization rotators with rotating polarizing beam splitters, which are then followed by two further receiver channels. The twin-channel Bell Test architecture has smaller systematic uncertainties than the single-channel test. The single-channel Bell Test was initially proposed as a more practicable instrument, but with more sophisticated technology the twin-channel measurement with its superior performance is now the preferred option, forming the basis for almost all Bell Tests today. As with the single-channel test, measurements are made at the Bell Test angles. In this case four effective coincidence counts are formed from Eq. (21), which are then used to determine the so-called quantum correlation, usually designated by $E$ for each of the Bell Test angles, from which the Bell Test test-statistic $S$ is then formed as detailed in [12] and [47].

*4.3. System signal-to-noise ratios and integration times*

One of the main sources of noise comes from the amplifiers and the thermal (black body) radiation entering the receiver from the ambient temperature surroundings. This noise power in each channel, referred to the amplifier input is just $FkT_0B$, where $F$ is the amplifier noise factor, $T_0$ is ambient temperature ~290 K and $B$ is the bandwidth of the system. For an amplifier noise factor of 2.5 and a 10 GHz bandwidth the noise is ~ 71 pW.

The entangled photon power at the input to the radio frequency amplifier is $P_S/L$, where $L$ is the loss in the entangled photon flux arising from stimulated emission and absorption. Since the entangled photon flux from the YIG has been estimated at $3.56 \times 10^{-12}$ Watts and all front-end components in front of the waveguide horn can be designed to be low loss, it is reasonable to assume L would be at the most a factor of two, which gives the entangled power flux at the amplifier input as $1.78 \times 10^{-12}$ Watts, for 5 Watts of pump power over one square centimeter. It is therefore also assumed that the loss of signal due to absorption and stimulated emission would still permit sufficient information from the signal and idler to be coherently integrated for the Bell Test.

As the YIG parametric amplification will also amplify thermally generated photons which have random phases, these will constitute extra noise in the system, having a power level $\bar{n}P_S/L$ where $\bar{n}$ is the mean photon number of Eq.(1). Since this is estimated at 604 at 10 GHz, the power flux from parametrically amplified thermal photons is $604 \times 1.78 \times 10^{-12} = 1.08 \times 10^{-9}$ Watts.

Summarizing the above, the signal-to-noise on each of the signal and idler arms of the interferometer on the input to mixer-1 is given by

$$SNR_1 = \frac{P_S/L}{\bar{n}P_S/L + FkT_0B}, \quad (22)$$

which has a numerical value of $1.55 \times 10^{-3}$. Given that mixer one multiplies the signal and idler with their independent noises together, the signal-to-noise ratio output from mixer-1 is

$$SNR_2 = \left(\frac{P_s/L}{\overline{n}P_s/L + FkT_0B}\right)^2, \qquad (23)$$

which has a numerical value of $2.41 \times 10^{-6}$.

The final stage of the receiver mixes the combined and signal and idler signatures (together with noise) down to a base-band frequency by mixing with the pump radiation in mixer-2. Coherent integration of the output from mixer-2 brings an improvement in the signal-to-noise ratio proportional to the root of the number of samples. Sampling at the Nyquist rate of 2B, the signal-to-noise ratio output after an integration time of $t_{INT}$ is

$$SNR_{OUT} = \left(\frac{P_s/L}{\overline{n}P_s/L + FkT_0B}\right)^2 \sqrt{2Bt_{INT}}. \qquad (24)$$

It can be seen from this that the highest signal-to-noise ratios will be achieved by using the largest possible radiation bandwidth. Inserting numerical values into this equation gives a signal-to-noise ratio of unity for an integration time of 8.59 s. Measurement of this signature at each of the Bell Test angles will therefore take a few tens of seconds.

As the parametric gain is increased, more entangled photons will be created and with a sufficiently high gain the parametrically amplified thermal photon fluctuations may dominate the noise generated by the RF amplifier. In this case $\overline{n}P_s/L \gg FkT_0B$ and Eq. (24) simplifies to

$$SNR_{OUT} = \left(\frac{1}{\overline{n}}\right)^2 \sqrt{2Bt_{INT}}. \qquad (25)$$

Transposing the last equation indicates the integration time to achieve a given signal-to-noise ratio is

$$t_{INT} = \frac{1}{2B} SNR_{OUT}^2 (\overline{n})^4. \qquad (26)$$

This indicates there would be benefits in moving up in frequency, so bandwidth B could be increased and the mean numbers of thermal photons would be reduced. Doubling the frequency of the pump from 20 GHz to 40 GHz (also requiring the doubling of $H_0$) and increasing the bandwidth from 10 GHz to 20 GHz, would shorten the integration time by a factor of 32.

*4.4. Closing Bell Test loopholes*

The two main loopholes in Bell Tests are the 'detector efficiency loophole' and the 'locality loophole', the former being associated with the fact that a photon counting detector does not ensure sufficient quantum efficiency and the latter associated with a local communication link between the two receivers which could corrupt the result.

For the case of the detection efficiency loophole this should be less of an issue with radio receivers, as information is more easily captured with homodyne quadrature receivers operating in the continuous-variable mode [12]. Noise is present due to the ambient thermal photons, which is removed by coherent integration, but the essential point is that the information about the arrival of each photon of an entangled pair is captured.

For the case of the locality loopholes, strategies to remove these would be the following. With the envisaged integration times being a matter of seconds and receivers placed several meters apart, it might be argued that there would be sufficient time for millimeter wave communication between receivers, potentially leading to corruption of data. However, this loophole would be closed by randomly changing the plane of polarization on a timescale sufficiently rapidly that measurement at one polarization is shorter than the time taken for signals to pass between the receivers and the source, as is done in the optical Bell Tests [48]. Rapidly switching the polarization can be done using phase shifters and waveplates constructed from pin or microwave varactor diodes, as these devices switch on nanosecond timescales. This switching would need to be done in synchronization with the data acquisition. This is entirely possible as the sample times required to satisfy the Nyquist criterion associated with GHz bandwidths is sub-nanosecond, a sampling capability readily available now.

In the case of the locality loophole associated with the link between the two receiver channels at the site of mixer-1 (before the sampling) this may be removed by sampling separately in each of the channels before mixer-1. The action of mixer-1 would then be performed digitally on uncorrupted samples, before a digitally downshifting action of mixer-2. In such a configuration it might be argued there may be a communication link via the sampling clock supplied to the two samplers. This link would be removed by having free-running sampling clocks at the sites of the samplers, which are only periodically up-dated to keep them in synchronization.

It is not being claimed here that this system offers a completely loophole free Bell Test, only that it offers a new potential direction for research into the phenomenon of entanglement, with attention paid to how loopholes may be closed. Furthermore, the approach here is only semiclassical and that a full quantum operator evaluation is necessary to identify potential quantum effects that may thwart the function of the proposed system. This would constitute an essential next phase of the work.

## 5. Conclusion

The proposed combination of a YIG parametric amplifier with a homodyne interferometer enables the measurement of a flux of entangled photons in the presence of a much larger flux of thermally generated photons. Calculations indicate that approximately 5 Watts of millimeter-wave radiation pumping 0.3 cm$^3$ of YIG will generate ~0.5 × 10$^{12}$ entangled photon pairs per second. The proposed architecture enables an ambient temperature Bell Test to be made which would require only tens of seconds of integration time. It is encouraging with this architecture of Bell Test that there are good opportunities to close the most severe loopholes of detector efficiency and locality. A successful demonstration of this test will lead to novel architectures of entanglement-based quantum technology systems, potentially for applications in computers and millimeter wave communication systems and sensors.

**Data Availability**

The datasets generated during and/or analyzed during the current study are available from the corresponding author on reasonable request.

**Acknowledgements**

The authors are grateful for the partial financial funding of this work from the TERANET Seedcord funding scheme provided by the UK EPSRC, through the School of Electronic and Electrical Engineering, at the University of Leeds, managed by Professor John Cunningham and Professor Martyn Chamberlain. The insightful comments and critique provided by the referees has been an enormous help in the refinement of the paper.

**Author contributions**

N.A.S. made the calculations of the entangled photon fluxes and the signal-to-noise ratios in the receiver systems. S.R.H. provided the details on the YIG ferrites, their hysteresis and use in resonance circuits. Both authors contributed to the manuscript.

**Competing Interests**

The authors declare that there are no competing interests.